%% file: ms.tex
\documentclass[twocolumn]{aastex63}
\usepackage{amssymb,graphicx,bm}
\usepackage{natbib}
\setcitestyle{notesep={ }} 
\newcommand{\dv}{\ensuremath{\Delta v_{\rm los}}}
\newcommand{\dproj}{\ensuremath{d_{\rm proj}}}

\newcommand{\arcs}{\ensuremath{^{\prime\prime}}}

\newcommand{\rev}[1]{{#1}} 
\newcommand{\revtwo}[1]{{#1}} 

\begin{document}

\title{A Very Rich Bimodal Galaxy Cluster Merger: RXC J0032.1+1808}

\shorttitle{A Very Rich Bimodal Galaxy Cluster Merger}

\author[0000-0002-0813-5888]{David Wittman}
\affiliation{Department of Physics and Astronomy, University of California, Davis, CA 
  95616 USA}
\author[0000-0002-6217-4861]{Rodrigo Stancioli}
\affiliation{Department of Physics and Astronomy, University of California, Davis, CA 
  95616 USA}
\author[0000-0001-8559-4538]{Christopher Hopp}
\affiliation{Department of Physics and Astronomy, University of California, Davis, CA 
  95616 USA}
\author[0009-0007-5074-5595]{Faik Bouhrik}
\affiliation{Department of Physics and Astronomy, University of California, Davis, CA 
  95616 USA}
\author[0000-0002-0587-1660]{Reinout van Weeren}\affiliation{Leiden Observatory, Leiden University, PO Box 9513, 2300 RA Leiden, The Netherlands}
\author[0000-0002-9325-1567]{Andrea Botteon}\affiliation{INAF - IRA, via P. Gobetti 101, I-40129 Bologna, Italy}

\keywords{Galaxy clusters (584); Galaxy
  spectroscopy (2171); Radio continuum emission(1340); Hubble Space
  Telescope (761)}

\begin{abstract} 
  The galaxy cluster RXC J0032.1+1808 has been
  well-studied with optical imaging and gravitational lensing mass
  maps, both of which reveal an elongated morphology in the
  north-south direction. We find that its X-ray morphology is bimodal,
  suggesting that it is in the process of merging; combined with a
  previously reported detection of a radio relic, we suggest that the
  system is seen after first pericenter. We extract the global X-ray
  temperature and unabsorbed luminosity from archival
  XMM-\textit{Newton} data, finding $T_X=8.5^{+1.1}_{-0.9}$ keV and
  $L_X=1.04 \pm 0.03 \times 10^{45}$ erg s$^{-1}$ at 90\% confidence
  in the $0.5$--$10.0$ keV energy range.  We conduct a redshift survey
  of member galaxies and find that the line-of-sight relative velocity
  between the two subclusters is \rev{1002–-1785 km/s at 68\%
    confidence, although the BCGs have an even larger difference of
    2466 km/s.}  We use publicly available hydrodynamic simulations to
  show that it cannot be a head-on merger; \rev{it is consistent with
    an impact parameter of 0.5 or 1 Mpc.  We also use the simulation
    set to find initial constraints on the time since pericenter
    (${\approx}315$--550 Myr) and the locus of possible lines of
    sight.}
\end{abstract}

\section{Introduction}\label{sec-intro}

Galaxy clusters are the largest virialized structures in the cosmos.
The merging of two such clusters releases up to $10^{64}$ erg of
energy that can drive shocks, heat the intracluster medium (ICM) gas,
and accelerate cosmic rays \citep[see][for a review]{Molnar16review}.
Cluster mergers can also probe dark matter (DM) properties
\citep[e.g.][]{Markevitch04,Clowe06,Bradac08-MACS0025,
  Randall2008,Dawson11, Jee15CIZA,A56}. A challenge in modeling
mergers is that they take ${\sim}1$ Gyr from first to second pericenter
passage, so each merging system is observed at only one instant in its
evolution.  Interactions of three or more bodies are thus difficult to
disentangle and trace back in time.  This motivates ongoing efforts to
find mergers that can be modeled as binary based on their galaxy
distribution in optical imaging surveys \citep{WHY2024,HoppXSORTER}.

One such effort is X-SORTER (X-ray Survey Of meRging clusTErs in
Redmapper), which selects binary merger candidates based on brightest
cluster galaxy (BCG) information in the redMaPPer
\citep{Rykoff2014,Rykoff2016} cluster catalog based on Sloan Digital
Sky Survey \citep[SDSS;][]{SDSS2000} imaging. The initial selection
criterion was that the top BCG candidates in a cluster are separated
by ${\ge}1\arcmin$; see \citet{HoppXSORTER} for more details. These
candidates are then subjected to a manual inspection of their X-ray
surface brightness (XSB) distribution relative to the BCGs. A cluster
with a single XSB peak between the BCGs is likely seen soon after a
close first pericenter passage, and is considered most interesting for
further followup.  Post-pericenter mergers discovered this way include
two using XSB data from the XMM-\textit{Newton} archive, Abell 56
\citep{A56} and RM J150822.0+575515.2 \citep{RMJ1508}; one using XSB
data from the \textit{Chandra} archive, the Champagne cluster
\citep{Champagne2025}; and a few candidates from a recent
XMM-\textit{Newton} survey of 12 systems with no prior archival data
\citep{HoppXSORTER}.

This paper highlights a system with widely separated BCGs \textit{and}
widely separated XSB peaks: RXC J0032.1+1808. A radio relic associated
with this system was identified by \citet{Botteon2022} in
the LOFAR Two-Meter Sky Survey
\citep[LoTSS-DR2;][]{LOTSS-Shimwell2022}, indicating that it is a
post-pericenter merger.  The lack of BCG-XSB separation then implies
that the merger was not head-on, and thus provides an interesting
counterpoint to the head-on examples cited above.

RXC J0032.1+1808 has been well studied with HST as part of the RELICS
project \citep{RELICS2019,Acebron2020}, but it has not been previously
recognized as a post-pericenter merger.  While making extensive use of
archival data, this paper also describes a new Keck/DEIMOS galaxy
redshift survey of the system. We combine this with archival redshifts
to measure the relative line-of-sight (LOS) velocity of the
subclusters.  We then use the mass peak separation, X-ray morphology
and temperature, relic position, and subcluster velocities to select
plausible merger models from a library of hydrodynamic simulations.

In \S\ref{sec-litrev} we provide an overview of the properties of RXC
J0032.1+1808 based on archival data; in \S\ref{sec-z} we describe the
redshift survey and subcluster velocity results; in \S\ref{sec-mm} we
derive merger models; and in \S\ref{sec-discussion} we conclude with a
brief discussion. We assume a flat $\Lambda$CDM cosmology with
$H_0=69.6$ km/s and $\Omega_m=0.286$. At the cluster redshift of
0.3732 (derived in \S\ref{sec-z}) the angular scale is thus 5.196
kpc/arcsec \citep{Wright2006CosmologyCalculator}.

\section{Cluster Overview Using Archival Data}\label{sec-litrev}

This cluster was first identified at visible wavelengths as ZwCl
0029.5+1750 \citep{Zwicky1965book} and in the X-ray as RXC
J0032.1+1808 in the Rosat All-Sky Survey \citep{Bohringer2000}.  Note
that the former used B1950 coordinates in the name while the latter
used J2000 coordinates. The redMaPPer \citep{Rykoff2014} optical
cluster catalog, as updated by \citet{Rykoff2016}, designated the
system more precisely as RM J003208.2+180625.3 and estimated a
photometric redshift of 0.398.  They also found an optical richness of
248, making it the second-richest cluster in the entire catalog of
${\approx}$26,000 clusters. According to the lensing-calibrated
mass-richness relation of \citet{RMmassrichnessrelation}, this
predicts a remarkably high mass of
$M_{200}=2.5 ^{+2.0}_{-1.1}\times10^{15}\ h^{-1}$ M$_\odot$.

Subsequent to redMaPPer, the cluster was identified via the
Sunyaev-Zel'dovich effect (SZE) as PSZ1 G116.48-44.47 \citep{PSZ1} and
PSZ2 G116.50-44.47 \citep{PSZ2}. The latter paper inferred
$M_{500}=7.61^{+0.57}_{-0.63} \times10^{14}$ M$_\odot$ from the SZE. Note
that $M_{\Delta}$ is the mass within a radius within which the mean
mass density is $\Delta$ times the critical density of the universe at
that redshift, so $M_{200}$ is expected to be ${\approx}40\%$ larger
than $M_{500}$ \citep{White2001}.  Hence the SZE result implies
$M_{200}\approx1.1\pm0.1 \times10^{15}$ M$_\odot$, which is consistent
with the richness proxy.

Most recently, the system was intensively studied at visible
wavelengths by the RELICS project \citep{RELICS2019}, who used the
name RXC J0032.1+1808. We adopt the same name for consistency with
that project.

\begin{figure*}
\centerline{\includegraphics[width=6in]{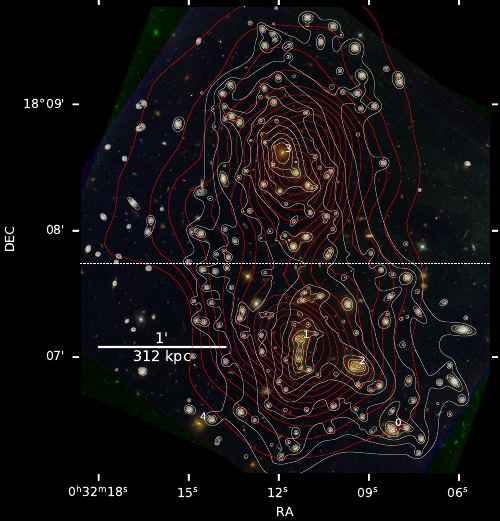}}
\caption{True-color HST/ACS image using the F435W, F606W, and F814W
  filters, with linearly spaced, point source subtracted X-ray surface
  brightness contours (red) and RELICS/GLAFIC convergence contours
  (white).  Numbers denote redMaPPer BCG candidates. The dashed line
  indicates the boundary between north and south subclusters used in
  \S\ref{sec-z}.}
\label{fig-relicsACS}
\end{figure*}

Figure~\ref{fig-relicsACS} presents a true-color HST/ACS image
released by the RELICS project using the F435W, F606W, and F814W
filters.\footnote{Available from
  \url{https://archive.stsci.edu/hlsps/relics/rxc0032p18/color_images/hlsp_relics_hst_acs_rxc0032p18_multi_v1_color.fits}}
We overplot the point source-subtracted X-ray surface brightness (red contours)
from the archival XMM-\textit{Newton} data described below. There are two
X-ray peaks separated by 1.27\arcmin\ (397 kpc) with a position angle
very close to north. Each X-ray peak is associated with one or more of
the five BCG candidates identified by redMaPPer (labeled 0-4 in
Figure~\ref{fig-relicsACS}) along with other galaxies of similar
color, with a dearth of galaxies in the central region between the
X-ray peaks.  The dashed line is drawn where the X-ray contours are
narrowest (at a declination of 18.1278 degrees), marking a nominal
division between north (N) and south (S) subclusters.  The SDSS
spectroscopic
database\footnote{\url{https://dr18.sdss.org/optical/spectrum/search}}
lists nine galaxies in the footprint of Figure~\ref{fig-relicsACS} at
$z{\approx}0.37$, confirming that both N and S are at the same
redshift.

In conjunction with the RELICS project, \citet{Acebron2020} used
strong gravitational lensing mass modeling to produce maps\footnote{Available from
  \url{https://archive.stsci.edu/hlsps/relics/rxc0032p18/models/}.} of
the convergence (a scaled surface mass density) $\kappa$  used later
in this paper.  They compared maps produced by four different modeling
methods, which all agreed on mass peaks coincident with BCG 3 in the
north and BCG 1 in the south. There was less agreement on additional
mass peaks in the south: all maps showed a peak at BCG 2, but three
maps showed it as a minor peak and one showed it nearly rivaling the
main southern peak at BCG 1. One map showed an additional peak in the
center, with the other three showing no sign of such a peak.  We
select the map produced by the GLAFIC code as the `median' map in
terms of these features, and plot it with white contours in
Figure~\ref{fig-relicsACS}. (The one aspect in which the GLAFIC map
stands out is the large amount of mass it assigns to individual member
galaxies, which are highly visible in Figure~\ref{fig-relicsACS}.)  We
add up the mass elements in the pixelized maps using the dashed line
in Figure~\ref{fig-relicsACS} as the boundary between north and south,
and find the north/south mass ratio is $1.06\pm0.07$.  This is a key
observation not available from the optical or SZE studies, which use
proxies for the total mass.

\textit{X-ray data.} The cluster was observed with the
XMM-\textit{Newton} European Photon Imaging Camera (EPIC) in 2010
(Obs.ID 0650380301, P.I. Allen). We reduced the X-ray data with the
XMM-\textit{Newton} Science Analysis System (\texttt{SAS}) version
19.0.0. After filtering out soft-proton flares, we ended up with an
effective exposure time of 6525 s, 6533 s, and 4373 s for the MOS1,
MOS2, and PN detectors, respectively. Point-source detection was
performed with the \texttt{SAS} task \texttt{edetect\_chain}, and
sources with a low likelihood of being extended were masked out. We
only considered single-to-quadruple events from MOS and
single-to-double events from the PN.

In order to obtain temperature and luminosity estimates for the
cluster, we defined a circular region with a 90\arcsec\ radius centered
between the two subclusters. We subtracted the background using the
double-subtraction method described by \citet{Arnaud2002}, which uses
both a blank-sky event list \citep{Carter2007} built from stacked,
source-removed, archival EPIC observations and an off-source region of
the current observation to account for the different spatial
dependence of cosmic and non-cosmic X-ray backgrounds; for a detailed
description of the method, we refer to \citet{Arnaud2002}. We chose a
110'' circular region to the northwest of the cluster with no
noticeable X-ray sources as our background region. Vignetting
correction was performed using the \texttt{evigweight} task. The
resulting spectra for the three EPIC instruments were simultaneously
fitted using \textsc{XSPEC} \citep{XSPEC}. We used an \texttt{apec}
model for the cluster emission multiplied by a \texttt{phabs} model to
account for the galactic extinction; the redshift was kept fixed at
$z=0.3732$ and the HI column density was set to $4.07 \times 10^{20} $
cm$^{-2}$ (obtained with the \texttt{FTOOLS} task \texttt{nH} using HI
maps from \citet{HI4PI}). Our best fit resulted in a global
temperature of $T_X=8.7^{+1.1}_{-0.9}$ keV and an unabsorbed
luminosity of $L_X=1.02 \pm 0.03 \times 10^{45}$ erg s$^{-1}$ in the
$0.5$--$10.0$ keV energy range, where the quoted uncertainties
represent the 90\% confidence interval.

Finally, we used the \texttt{ESAS} package and followed the
prescription in the ESAS Cookbook \citep{ESAS} to obtain an
exposure-corrected, background-subtracted image in the 0.40-1.25 keV
energy band, adaptively smoothed with the \texttt{smoothingcounts}
parameter set to 50. To obtain the contours in
Figure~\ref{fig-relicsACS}, we applied an additional 6.75\arcs\
smoothing (roughly the FWHM of the PN camera).

\textit{Radio data.} \citet{Botteon2022} conducted a comprehensive
search for diffuse emission in LoTSS-DR2. They found two distinct
diffuse sources associated with this cluster, one near the center and
one on the northern outskirts.  The central emission is classified as
a radio halo, with the highest surface brightness centered on the
northern subcluster. On the northern outskirts approximately 650 kpc
from the northern XSB peak and 940 kpc from the system center, an
elongated radio source was detected, with a largest linear extent of
550~kpc. \citet{Botteon2022} classified this source as a radio
relic. Radio relics are typically extended sources found on the
outskirts of disturbed clusters, believed to trace shocks; for a
review, see \citet{ReinoutRadioReview19}.

\begin{figure*}
\centerline{\includegraphics[width=5in]{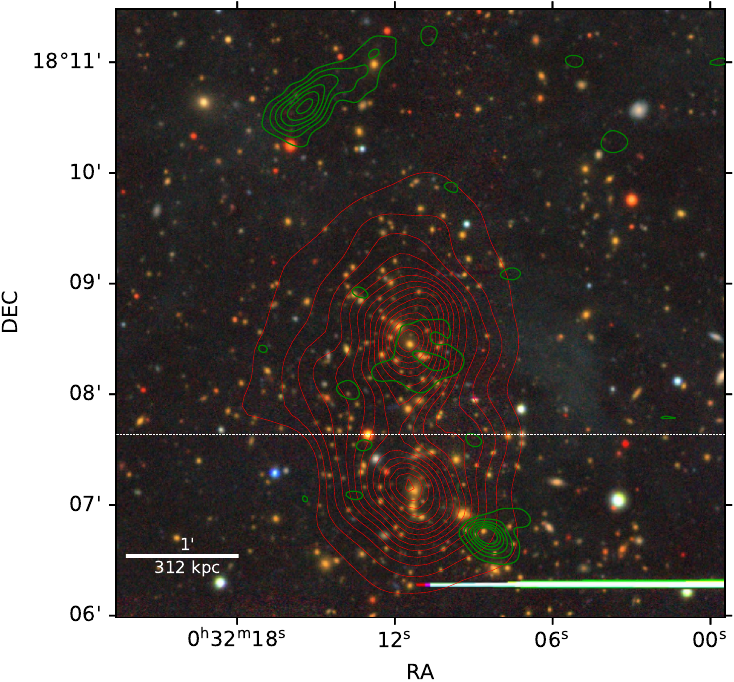}}
\caption{Linearly spaced LoTSS (green) and XMM-\textit{Newton}
  contours (red) over a true-color $griz$ image from the DESI Legacy
  Survey. The radio relic likely traces a shock, suggesting that the
  two subclusters have already experienced a pericenter passage.  This
  field is larger than that of Figure~\ref{fig-relicsACS},
  illustrating that the relic is rather far from the system center
  compared to the subclusters. The dashed line is the same as in
  Figure~\ref{fig-relicsACS}. The artifact at lower right is from a
  7th magnitude star outside the region shown.}
\label{fig-legacy}
\end{figure*}

The relic detected in this system falls outside the ACS field of
view. Consequently, Figure~\ref{fig-legacy} presents a broader field,
incorporating LoTSS contours in green obtained from the LOFAR
source-subtracted image, with a resolution of
$14 \arcsec \times 7 \arcsec$, released by \citet{Botteon2022}. The
color image in Figure 2 is a $griz$ image from the DESI Legacy Survey
DR10\footnote{\url{https://www.legacysurvey.org/dr10/description/}},
and the X-ray contours are identical to those in
Figure~\ref{fig-relicsACS}. The relic has an integrated 150 MHz flux
density of $13.00 \pm 1.54$ mJy, corresponding to a 150 MHz radio
power of $(7.2 \pm 0.8) \times 10^{24}$ W Hz$^{-1}$. The nature of the
diffuse emission remains to be confirmed with polarization and/or
spectral index measurements. Other potential origins of this extended
source, such as an active galactic nucleus (AGN), may need
consideration, although they are less likely. Notably, no optical
counterpart is observed at the extremities of the source, discounting
the possibility that the emission originates from a head-tail radio
galaxy. The elongation of the radio source with respect to that of the
X-ray emission is consistent with what is expected for a relic
source, generally appearing elongated perpendicular to the thermal gas
elongation. Moreover, the outer edge of the source is more sharply
defined, while the emission diminishes more gradually towards the
cluster center. Hence, we proceed with the assumption that the radio
source indeed traces a merger shock.

Without the LoTSS data, the lack of displacement between X-rays and
galaxies could have been taken as suggestive of a pre-pericenter
system. However, the association between relics and shocks strongly
suggests that this is a post-pericenter system \citep[e.g.,][]{WonkiTNGRelics2024,WonkiTNGRelics2025}.  Furthermore, the
relic has a position and morphology well matched to a
pericenter passage in the N-S direction suggested by the current
subcluster separation vector.  Assuming there has been a pericenter
passage, the lack of displacement between X-rays and
galaxies then suggests that the collision was not head-on. We support this
conjecture in more detail in \S\ref{sec-mm}.

\section{Redshift survey}\label{sec-z}

The goals in this section are to: (i) identify any
foreground/background structures that may affect the interpretation of
X-ray and lensing observations; and (ii) collect redshifts of cluster
members to analyze the LOS velocity structure of the
system. In the absence of clear LOS velocity features, task (ii)
reduces to quantifying the LOS velocity of the northern subcluster
relative to the southern subcluster.

\subsection{Observations}
\textit{Observational setup.} We observed RXC J0032.1+1808 with the
DEIMOS multi-object spectrograph \citep{FaberDEIMOS} at the W. M. Keck
Observatory on July 1, 2022 (UT). The DEIMOS field of view is
approximately 16$^\prime \times 4^\prime$, easily encompassing the two
subclusters visible in Figures~\ref{fig-relicsACS}
and~\ref{fig-legacy} as well as potential foreground/background
structures.  We used the slitmask design software \texttt{dsimulator}
to design a slitmask with 95 slits.  Given a list of potential targets
with associated priorities, where priority is a positive number,
\texttt{dsimulator} attempts to maximize the total priority of
assigned slits while preventing slit collisions. We used Pan-STARRS
photometric redshifts \citep{PSphotoz2021} and photometry to calculate
a priority for each target. The initial priority is the likelihood of
each galaxy being at $z=0.38$; given the large photometric redshift
uncertainty (typically 0.16), this enhances the chance of targeting a
cluster member but still allows a broad selection of galaxies. We then
multiplied that priority by $(24-r)$, where $r$ is the apparent $r$
magnitude, to prioritize brighter galaxies more likely to yield a
secure redshift.  We removed galaxies with known redshift from the
target list to avoid redundancy, except for galaxies in the outer half
of the field where slit collisions would be rare. There, we allowed
for duplicate redshifts to assess repeatability.

We used the 1200 line mm$^{-1}$ grating, which results in a pixel
scale of 0.33 \AA\ pixel$^{-1}$ and a resolution of ${\sim}1$ \AA\ (50
km/s in the observed frame). The grating was tilted to observe the
wavelength range $\approx$ 4900--7500 \AA\ (the precise range depends
on the slit position), which at the cluster redshift includes spectral
features from the [OII] 3727 \AA\ doublet to well past the magnesium line at
5177 \AA. The total exposure time was 45 minutes,
divided into three exposures. The seeing was roughly 1\arcsec, with
minor variations over time.

\textit{Data reduction and redshift extraction.} We calibrated and
reduced the data to a series of 1-D spectra using PypeIt
\citep{pypeit:joss_pub,pypeit:zenodo}.  We double-checked the arc lamp
wavelength calibration against sky emission lines, and found good
agreement. To extract redshifts from the 1-D spectra we used custom
Python software described in more detail in \citet{A56}.  We found 43
secure redshifts, with a typical redshift uncertainty of 0.0001 (22
km/s in the frame of the cluster) or better.  These are listed in
Table~\ref{tab-zspec}.

\begin{deluxetable}{llll}
  \tablecaption{Galaxy redshifts}  \label{tab-zspec}
  \tablecolumns{4}
  \tablehead{\colhead{RA (deg)} & \colhead{DEC (deg)} & \colhead{z} & \colhead{uncertainty}}
  \startdata
\input{ztable.tex}
    \enddata
\end{deluxetable}

\textit{Archival redshifts.} We searched the SDSS redshift
database\footnote{\url{https://dr18.sdss.org/optical/spectrum/search}},
which provided 60 redshifts within a 10\arcmin\ radius.  We also
searched the NASA/IPAC Extragalactic Database \citep{NEDDOI}, but
found no `SLS' redshifts (a flag indicating a spectroscopic redshift
based on multiple lines) that were not in the SDSS database.  Of the
60 galaxies with archival redshifts, there was one duplicate with our
results, which matched with a redshift difference of 0.000155 or 34
km/s in the frame of the cluster.

\subsection{Subclustering and kinematics}\label{ssec-kinematics}

\begin{figure}
\centerline{\includegraphics[width=\columnwidth]{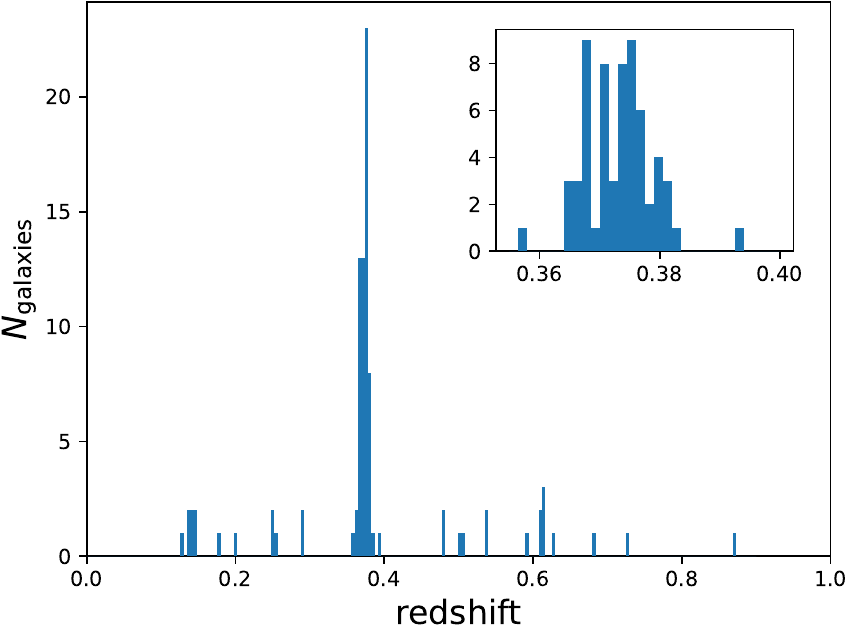}}
\caption{Redshift histogram, with inset showing the redshift interval 
  around RXC J0032.1+1808.}
\label{fig-zhist}
\end{figure}

The merged catalog contains 103 redshifts within a 10\arcmin\ radius.
Figure~\ref{fig-zhist} shows a histogram of these redshifts. While
probing a wide range of redshifts, these data reveal no notable
foreground/background structures. \rev{However we caution that
  Figure~\ref{fig-zhist} is not unbiased: its construction favored
  brighter galaxies, those in the photometric redshift range
  0.21--0.53, and/or those with more easily identifiable spectral
  features leading to a secure redshift. Hence it cannot be taken as
  definitively ruling out additional structures.} 

The inset in Figure~\ref{fig-zhist} illustrates 62 galaxies which
could be plausibly considered cluster members by virtue of being in
the redshift range 0.35--0.40. The galaxy at $z=0.3925$ is far outside
the cluster spatially, outside the DEIMOS footprint and at the very
edge of our generous 10\arcmin\ search radius for archival
redshifts. We therefore do not consider it a cluster member. We assess
the mean redshift and \rev{rest-frame} velocity dispersion of the
remaining 61 galaxies using the biweight estimator \citep{Beers1990}
to provide robustness against outliers.  With this estimator, the
inclusion or exclusion of the galaxy at $z=0.3574$ has only a minor
effect: $\bar{z} = 0.37325\pm0.00074$ (excluded)
vs. $0.37317\pm0.00070$ (included), and $\sigma_v=1091\pm 81$ vs
$1134\pm97$ km/s respectively (uncertainties are derived from
jackknife resampling).

\begin{figure}
\centerline{\includegraphics[width=\columnwidth]{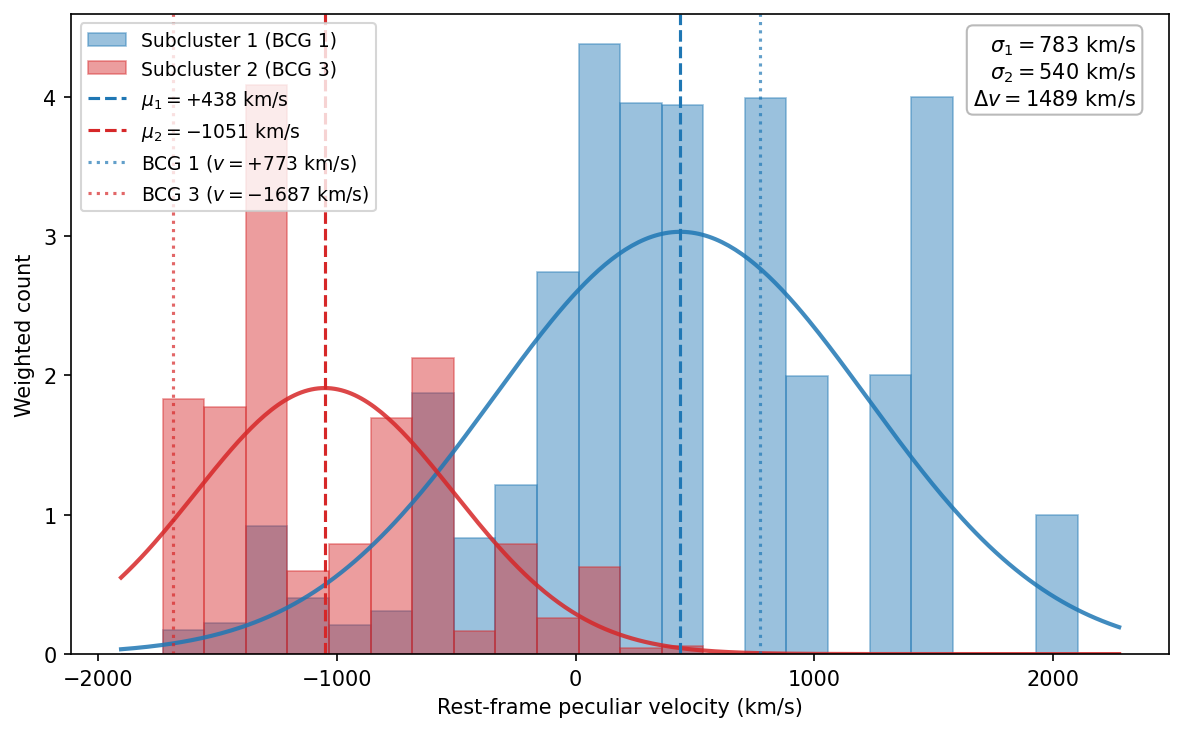}}
\caption{\revtwo{Results derived from a simultaneous fit of two
    subclusters (with King spatial and Gaussian velocity profiles) to
    328 photometric member positions and 51 spectroscopic member
    positions and redshifts. This visualization focuses on velocities,
    with the best-fit velocity distributions shown as curves, their
    peaks marked with dashed lines, and the BCG velocities marked with
    dotted lines. The best-fit models yield a fractional probability
    of each spectroscopic galaxy belonging to each subcluster; each
    galaxy contributes to the red (blue) histogram in proportion to
    its probability of membership in the northern (southern)
    subcluster.  BCG 3 has an extreme velocity even within its
    subcluster. Photometric members do not appear in this
    visualization, but did support the fit by defining the sky
    positions of the subclusters better than possible with the
    spectroscopic catalog alone. }}
\label{fig-vfit}
\end{figure}

An Anderson-Darling test demonstrates consistency ($p>0.15$) with a
single Gaussian redshift distribution in this range. \rev{At first
  glance this suggests a minimal line-of-sight velocity difference
  between the subclusters. (We use the symbol \dv\ to denote this
  velocity in the cluster frame.)  However, inferring \dv\ from the
  BCGs would lead to a different conclusion: the northern BCG (labeled
  3 in Figure~\ref{fig-relicsACS}) has $\dv=2460$ (3170) km/s relative
  to BCG 1 (2) in the south.  The tension between the all-members and
  BCG estimates of \dv\ is due to two possibly related factors. First,
  within the northern subcluster the BCG has an apparently atypical
  velocity, as shown by the red dotted line in
  Figure~\ref{fig-vfit}. Second, the small spectroscopic sample size
  hampers the all-members inference as follows.}

We created mock spectroscopic catalogs drawn from hypothetical
clusters with $\dv=2500$ km/s and compared two inference methods: (i)
averaging the redshifts on either side of a line in the sky as drawn
in Figure~\ref{fig-relicsACS}; and (ii) a Gaussian mixture model (GMM)
based on two sky coordinates as well as the redshift.  \revtwo{These
  mock catalogs were drawn from 3-D Gaussians with the same spatial
  and velocity dispersions as our spectroscopic catalog. With either
  inference method} the inferred \dv\ was close to zero---in other
words, highly biased--- with the relatively small catalog and \dproj\
observed here.  This is because overlap on the sky leads to mistaken
membership assignment which always reduces the inferred \dv.  Hence it
is possible that the BCGs are tracing a true $\dv{\approx}2500$ km/s
that is not evident in the \revtwo{current} spectroscopic catalog.
\revtwo{We then varied \dproj\ and found, as expected, that the bias
  in either method decreases as \dproj\ increases. We also found that
  the bias in the line-in-sky method depends only on the ratio of
  \dproj\ to subcluster size, $\frac{\dproj}{\sigma_{\rm proj}}$,
  regardless of sample size, because the fraction of members on the
  wrong side of the true bisector (the line we used) is fixed by
  $\frac{\dproj}{\sigma_{\rm proj}}$.  Small sample size, in contrast,
  exacerbated the misidentification problem for the GMM, because it
  adjusted its spatial centers to follow randomly realized density
  peaks.  Still, with sample sizes of hundreds, the GMM outperforms
  the line-in-sky method, and also reaches negligible bias when
  $\frac{\dproj}{\sigma_{\rm proj}}\approx 5$. This agrees with
  previous work on such systems \citep{MCCsampleanalysis}.  These
  conclusions remained true when we used \citet{King1972} spatial
  profiles, which are commonly used to fit the projected density of
  cluster member galaxies \citep{Bahcall1999}, with Gaussian velocity
  profiles.}
   
\revtwo{With $\frac{\dproj}{\sigma_{\rm proj}}\approx 1.5$ and 51
  spectroscopic members in this case, a new approach is required. We
  use photometric members to better support the spatial part of the
  fit as follows. Given a putative subcluster center at 2-D
sky position $\bm{r}$ and redshift $z$, the King profile
likelihood of the $i$th photometric galaxy at $\bm{r}_i$ is
\begin{equation}
  \label{eq:1}
  \mathcal{L}_{\rm phot} \propto \Sigma(1+\frac{|\bm{r}_i - \bm{r}|^2}{R_c^2})^{-1}
\end{equation}
where $\Sigma$ is the central surface density and $R_c$ is the core
radius. The likelihood of the $j$th spectroscopic galaxy with sky
position $\bm{r}_j$ and redshift $z_j$ is
\begin{equation}
  \label{eq:2}
  \mathcal{L}_{\rm spec} \propto
  e^{-\frac{(z_j-z)^2}{2\sigma_z^2}}\Sigma(1+\frac{|\bm{r}_j - \bm{r}|^2}{R_c^2})^{-1}
\end{equation}
To provide spatial coverage beyond the HST/ACS footprint, similar to
that of the 51 spectroscopic members, we selected 328 red-sequence
members (lacking redshifts) from Legacy Surveys DR10. We found the
maximum likelihood fit for two subclusters, A and B, with six free
parameters ($\bm{r},z,\sigma_z,\Sigma,R_c$) for each subcluster plus
one free parameter representing a uniform fore/background density.}
The fitted centers of the profiles \revtwo{landed} near, but not
exactly on, BCGs 3 and 1. \revtwo{Although the photometric members
  cannot contribute directly to the velocity estimate, they provide
  valuable spatial support for extracting subcluster velocities from
  the spectroscopic sample.  We confirmed this with mock datasets
  similar to RXC J0032.1+1808, finding a \dv\ bias far lower than the
  other two approaches: only 8 km/s (53 km/s) when the subclusters
  were truly separated by 1500 (2500) km/s.}

\revtwo{Figure~\ref{fig-vfit} shows the velocity dimension of our data
  and maximum-likelihood subcluster models. We find $\dv=1489$ km/s
  (dashed lines), which our mocks suggest is minimally biased.  We use
  Markov Chain Monte Carlo (MCMC) exploration of the likelihood to
  find 68\%, 90\%, and 95\% confidence intervals (CI) on \dv:
  1002--1785, 498--1891, and 257--1941 km/s respectively.  The BCG
  velocities are indicated with dotted lines in Figure~\ref{fig-vfit}:
  BCG 1 is $+335$ km/s from the peak of its (southern) subcluster
  velocity distribution, while BCG 3 is $-636$ km/s from its
  subcluster peak. This puts their velocity difference (2460 km/s)
  outside the 95\% CI provided by the full dataset.  To visualize
  this, we use the best-fit models to assign fractional memberships to
  each spectroscopic member. Each member then contributes partially to
  the blue (red) histogram for the southern (northern) subcluster in
  Figure~\ref{fig-vfit}. It is clear that BCG 3 is on the very edge of
  its subcluster velocity distribution. This is potentially worthy of
  further investigation with a larger spectroscopic sample, but for
  this work we adopt $\dv=1489$ km/s based on the full dataset.}

\section{Merger modeling}\label{sec-mm}
 
We compare the observations with the Parameter Space Exploration of Galaxy Cluster
Mergers\footnote{\url{http://gcmc.hub.yt/fiducial/index.html}}
\citep{ZuHone11,ZuHone2018} which is a suite of hydrodynamic, binary merger
simulations spanning a coarse grid of mass ratios (1:1, 1:3, and
1:10), impact parameters $b$ (0, 0.5, and 1 Mpc), and viewing angles
(along each coordinate axis; $x$ is parallel to the initial velocity
vector, $y$ defines the orbital plane with $x$, and $z$ is
perpendicular to the orbital plane).  The simulated primary cluster
mass is $M_{200} = 6\times10^{14}$ M$_\odot$, yielding a total of
$1.2\times10^{15}$ M$_\odot$ for an equal-mass merger, in good
agreement with the mass estimates in \S\ref{sec-litrev}.

The simulation products are 2-D maps of mass density, X-ray emissivity
(i.e., surface brightness in this 2-D context), ICM temperature, and
ICM pressure as quantified by the Compton $y$-parameter.  Given the
lensing convergence maps, we focus on the 1:1 mass ratio data
products.  We note that the `initial' (when the subcluster centers are
separated by $2r_{200}$) velocity in the simulations was set to 1200
km/s and not varied. Changing the initial velocity would change the
mapping from time since pericenter (TSP) to subcluster separation and
velocity. \rev{Similarly, the initial plasma conditions were fixed in
  this simulation suite; changing them would change $T_X$ as a
  function of time.} Hence the \rev{model parameters} derived below
are intended as a rough guide rather than a rigorous constraint.
Italics are used to introduce each observable.

\textit{XSB morphology.} There are two distinct XSB peaks. This is not
seen in any $b=0$ snapshot after pericenter, regardless of viewing
angle.  We therefore rule out $b{=}0$.  The remaining options, $b{=}0.5$
and $b{=}1$ Mpc, \rev{retain distinct XSB peaks closely attached to
  the DM peaks, as in the data, until TSP${\approx}1.4$ Gyr, when the
  impending second pericenter passage erases the distinct peaks.
  In summary, only $b{>}0$ and TSP\,$\lesssim1.4$ Gyr remain viable
  given the XSB morphology.}

\textit{Relic position.} In magnetohydrodynamic simulations, the shock
distance from the system center increases monotonically with time
\citep[e.g, ][]{WonkiTNGRelics2024,WonkiTNGRelics2025}, making it a
good clock for measuring TSP. In the \citet{ZuHone11} simulation
products, we tabulate the shock distance from the system center by
locating the maximum ICM temperature gradient in each
snapshot.\footnote{In one snapshot the maximum gradient fell on an
  interior feature and we manually marked the highest gradient
  associated with the outwardly propagating shock.}  The shock becomes
clear around 200 Myr after pericenter and moves rapidly, surpassing
the observed projected distance from the system center at 315 Myr
\rev{(we neglect the small $b$ dependence of these numbers for
  simplicity of presentation)}.  A relic can appear projected at a
smaller radius but not a larger one \citep{Skillman13}, yielding a
one-sided constraint that TSP $\gtrsim 315$ Myr.
  
\begin{figure}
  \centerline{\includegraphics[width=\columnwidth]{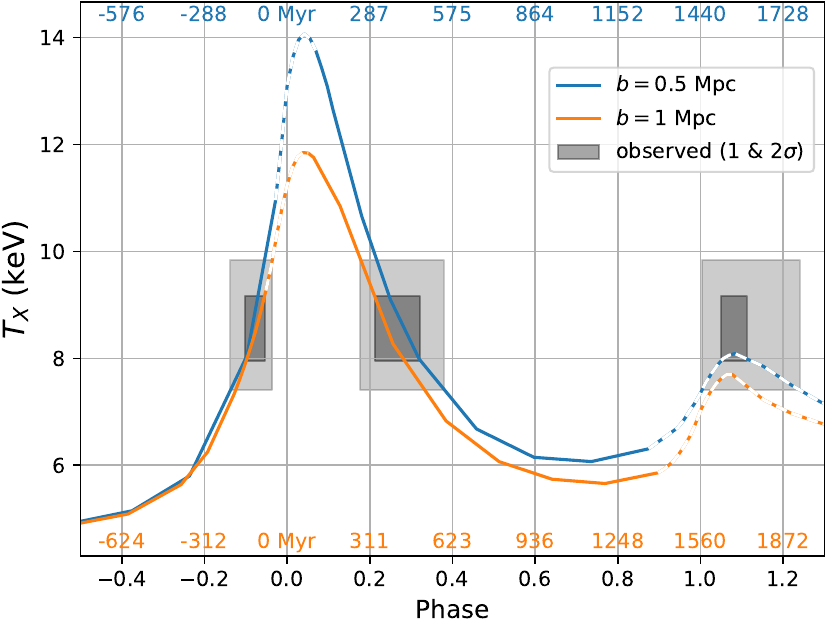}}
  \caption{Simulated global $T_X$ as a function of \rev{orbital phase
      (defined as zero at first pericenter and one at second
      pericenter), along with the observational constraints. The
      dashed parts of the curves indicate phases ruled out by XSB
        morphology, with the
        simulations lacking two distinct XSB peaks as observed.
      The pre-pericenter window is viable based on XSB and $T_X$, but
      inconsistent with the existence of the radio relic. Phases in
      the range 0.18--0.38 remain viable. Time since pericenter (in
      Myr) is marked along the top (bottom) edge for the hotter
      (cooler) $b=0.5$ (1) Mpc scenario.}}
\label{fig-TX}
\end{figure}

\textit{Global ICM temperature.} \S\ref{sec-litrev} found this to be
$T_X=8.5^{+1.1}_{-0.9}$ keV at 90\% confidence. To obtain a global
temperature from the simulated map, we compute the emission-weighted
temperature over the entire $7\times7$ Mpc box. Although this is
larger than the region over which $T_X$ is measured, we found that the
simulated global $T_X$ is insensitive to box size because the
emissivity is extremely low in the outskirts. We also confirmed that
the $T_X$ extraction is insensitive to the projection used. In
Figure~\ref{fig-TX} we plot the $T_X$ evolution of the simulations
\rev{versus orbital phase (defined here as zero at first pericenter
  and one at second pericenter), along with the observed 1 and
  $2\sigma$ ranges.  The observed $T_X$ is consistent with a brief
  window before first pericenter, a longer window between first
  pericenter and apocenter, and a third window around second
  apocenter.  The first window takes place before a shock is launched,
  so it can be ruled out based on the observed existence of a relic.
  The third window is ruled out based on XSB morphology: the dashed
  parts of the curves indicate times where the simulated XSB peaks are
  separated by less than the observed \textit{projected} separation
  (including snapshots lacking two distinct XSB peaks in any of the
  three projections). Hence, the only viable window is at phases
  0.22--0.38 (320--550 Myr after first pericenter) for $b=0.5$ Mpc,
  and phases 0.18-0.33 (280--520 Myr) for $b=1$ Mpc. These ranges
  almost entirely agree with the conclusion from the relic that TSP
  $\gtrsim 315$ Myr.}

\textit{Subcluster projected separation \dproj.}  We find this to be
464 kpc in the GLAFIC map and 446 kpc for the BCG 3--BCG 1 separation.
We adopt the former but mention the latter for robustness, as the
GLAFIC map has many alternate peaks. The adopted value does not
substantially affect the conclusions that follow. Note that \dproj\ is
sensitive to TSP as well as the viewing angle. \rev{Without a viewing
  angle constraint, the \dproj\ datum only rules out
  timesteps where the 3-D subcluster separation in the model is $<464$
  kpc, or} from about 110 Myr before pericenter
to 120 Myr after \rev{(phases ${\approx\pm0.1}$), which were already
  ruled out by other considerations.}

\begin{figure}
  \centerline{\includegraphics[width=\columnwidth]{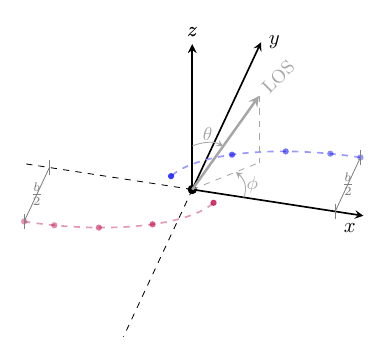}}
  \caption{\rev{Coordinate systems used in this section. The
      simulations use a Cartesian coordinate system with the orbit
      (stylized here as a motion diagram) in the $xy$ plane. The initial
      value of $y$ defines the impact parameter $b$, as the initial
      velocity is in the $x$ direction. We define a
      LOS with a spherical coordinate system using a colatitude
      $\theta$ and azimuthal angle $\phi$ which increases from $x$
      toward $y$.} }
\label{fig-coordsys}
\end{figure}

\textit{Subcluster relative LOS velocity \dv.} \rev{This observable
  complements \dproj: placing a lower bound on the 3-D speed rules out
  phases near \textit{apocenter} where speeds are low.  Due to its
  large uncertainty (\S\ref{ssec-kinematics}), \dv\ alone does not
  place strong constraints on TSP and/or viewing angle, so we use it
  in combination with \dproj.  We define viewing angle coordinates
  $(\theta,\phi)$ as illustrated in Figure~\ref{fig-coordsys}:
  $\theta$ is a colatitude away from the $z$ axis (which is defined in
  the simulations as perpendicular to the orbital plane), and $\phi$
  is an azimuthal angle from $x$ (which is the direction of the
  initial velocity) toward $y$ (which defines the impact
  parameter). Due to the symmetry, we consider only the spherical
  quadrant from 0--90$^\circ$ in $\theta$ and $\pm90^\circ$ in
  $\phi$.}

\begin{figure*}
  \centerline{\includegraphics[width=2\columnwidth]{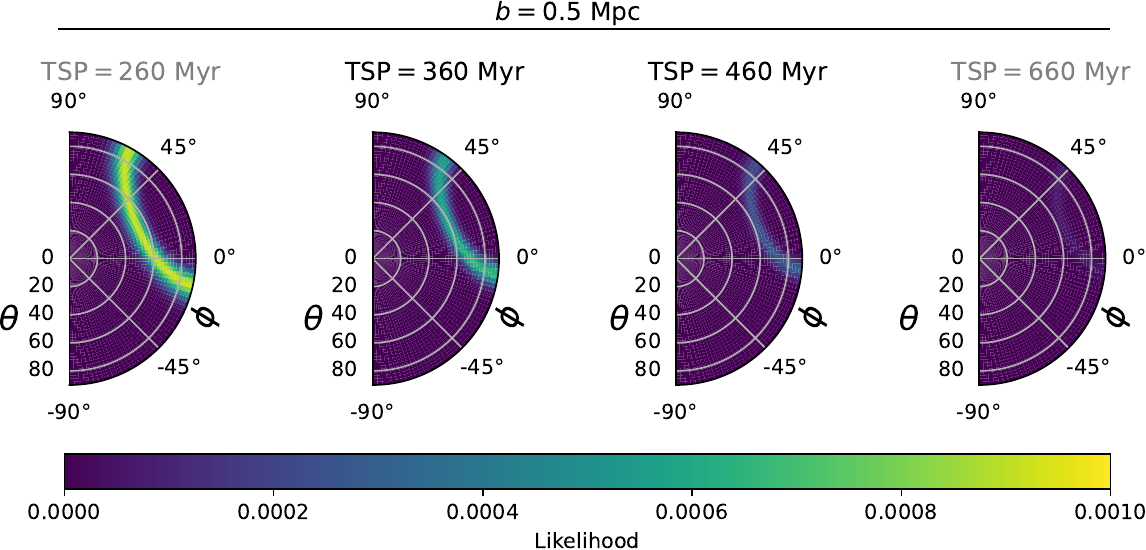}}\vskip16pt
    \centerline{\includegraphics[width=1.5\columnwidth]{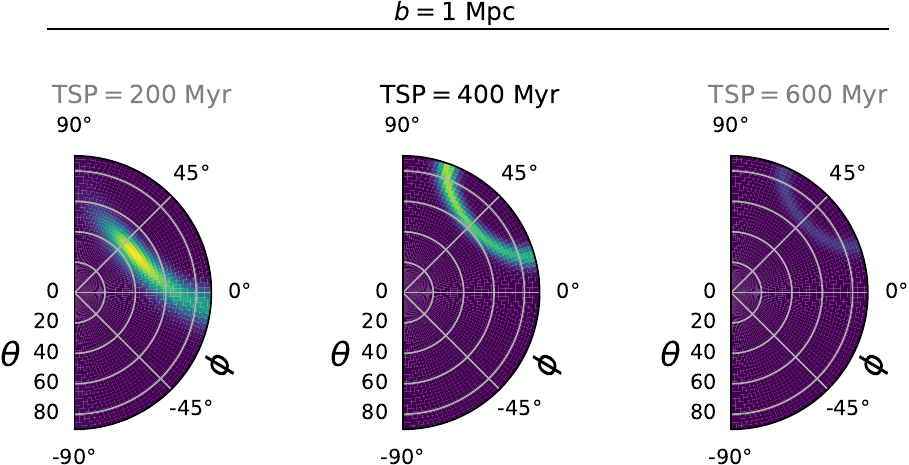}}
    \caption{LOS likelihoods based on the kinematic observables
        \dv\ and \dproj.  \textit{Top:} for $b=0.5$. The middle
        timesteps are allowed by $T_X$, while those with gray titles
        are too hot or too cold by ${>}2\sigma$ but are shown for
        context.  The two types of observables are complementary in
        the sense that they favor different but overlapping TSP
        ranges.  The locus of allowed LOS is \revtwo{largely static
          because the orbit is nearly radial at these times. The
          separation vector points to $\theta=90^\circ$ (by definition
          of the orbital plane) and $\phi\approx23^\circ$ (by
          happenstance). LOS $\approx30^\circ$ from there foreshorten
          that vector to match the observed \dproj\ while retaining a
          large fraction of the velocity along the LOS.}
        \textit{Bottom:} for $b=1$, which has coarser time sampling at
        the relevant times, and \revtwo{a separation vector that
          points to $\phi\approx45^\circ$.}  The likelihood values are
          similar, indicating that either value of $b$ produces viable
          models. \revtwo{The complementarity of X-ray and kinematic
        observables, and the stability of the allowed LOS locus,
        remain as noted for the $b=0.5$ scenario.}}
\label{fig-Lmap}
\end{figure*}

\rev{For each LOS in this spherical quadrant we compute a likelihood
  as follows. We identify the $x$ and $y$ position for each mass
  component at each snapshot and fit cubic splines for $x(t)$ and
  $y(t)$, then take time derivatives to find $v_x(t)$ and $v_y(t)$. We
  then dot that model velocity vector with the LOS to find the model
  prediction for \dv, and compute the likelihood from the MCMC
  exploration of the redshift survey (supported by the photometry to
  help determine the subclusters, as described in
  \S\ref{ssec-kinematics}).  We also compute the model \dproj\
  (between mass peaks) and compute the likelihood using a 50 kpc
  Gaussian uncertainty. The product of these two likelihoods is
  depicted in Figure~\ref{fig-Lmap}. The maps
  bracket the TSP window allowed by $T_X$; timesteps that are too hot
  or too cold by ${>}2\sigma$ have gray titles.  The time evolution of
  the LOS likelihood (for either value of $b$) illustrates the
  complementarity of the kinematic and X-ray observables: the latter
  are more effective at ruling out the earlier times shown in these
  figures. Strikingly, the allowed LOS form a well-defined locus that
  is roughly the same for either value of $b$, and persistent over
  time.  The likelihood of this locus fades toward the later
  timesteps, so 360 \revtwo{(400)} Myr is the best-matching TSP
  \revtwo{for $b=0.5$ (1)} considering both
  types of observables.  }

In summary, the staged simulations point to:
\begin{itemize}
\item a substantial impact
parameter. Although $b=0$ can be ruled out, there is remarkably little
difference between $b=0.5$ and $b=1$ for the observables at
hand. Sampling $b$ between 0 and 0.5 Mpc with additional
simulations would be useful for establishing a more precise lower
limit on $b$.  
\item \rev{TSP of 320--550 (315--520) Myr for $b=0.5$ (1). For the
    former, both lower and upper bounds are set by $T_X$. For the
    latter, the lower bound is from the relic position and is just 35
    Myr above the lower bound from $T_X$. The kinematic observables
    \dv\ and \dproj\ show a higher likelihood toward earlier parts of
    these windows, but do not conclusively narrow the window.}
\item \rev{viable LOS covering a fairly specific locus in parameter
    space as illustrated by Figure~\ref{fig-Lmap}. This locus allows for a substantial
    $\dv\approx1500$ km/s while \revtwo{foreshortening the separation vector
    by a factor $\approx2$ to maintain $\dproj\approx 450$ kpc.}}
\end{itemize}
\rev{These constraints represent only a first attempt at modeling the
  system with readily available tools, and are subject to caveats and
  limitations as discussed below.}

\section{Summary and Discussion}\label{sec-discussion}

RXC J0032.1+1808 has been well studied at particular wavelengths, but
we argue that its true nature has not been recognized: it is a
post-pericenter major merger. With distinct XSB peaks, it could be
mistaken for a pre-pericenter system but for the radio relic. \rev{By
  comparison with a public set of hydrodynamical simulations, we find
  the XSB morphology (two peaks with a minimum separation given by the
  observed separation) consistent with a substantial impact parameter
  ($b=0.5$ or 1 Mpc) and a range of times up to $1.4$ Gyr after first
  pericenter.  Requiring a match to the global $T_X$ substantially
  narrowed the viable time window, to 320--550 (280--520) Myr for
  $b=0.5$ (1). Including the relic position as a proxy for TSP
  narrowed the constraint slightly, to 315--520 Myr for $b=1$.}

We conducted a Keck/Deimos galaxy redshift survey of the area \rev{but
  the rest-frame LOS velocity difference between the subclusters
  remains uncertain due to substantial overlap of the subclusters on
  the sky. Formally, we find $\dv=1489^{+296}_{-487}$ km/s at 68\%
  confidence using 61 members, but this is in some tension with the
  2460 (3170) km/s between BCG 3 in the north and 1 (2) in the south.
  A much larger spectroscopic survey would be needed to fully probe
  the velocity structure of the system.  We find that any of the
  snapshots in the TSP windows cited above include LOS that are
  consistent with the kinematic data (the all-member $\dv$ constraint
  combined with the projected separation of the masses). The kinematic data slightly
  prefer times earlier in those windows, but are more useful at
  constraining the LOS than constraining TSP.}

\begin{figure}
  \centerline{\includegraphics[width=\columnwidth]{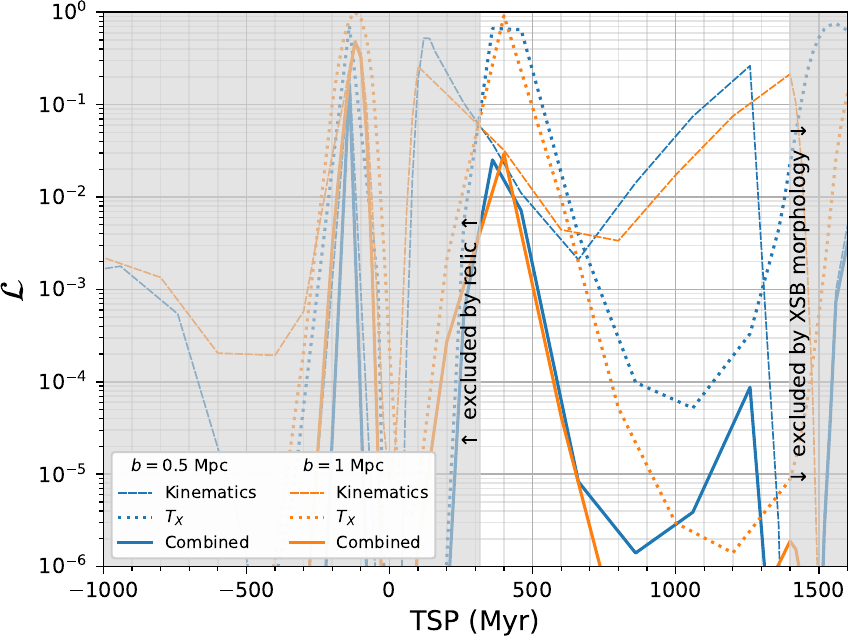}}
  \caption{\rev{Summary of likelihood $\mathcal{L}$ versus TSP,
      calculated based on kinematics ($\dv$ and $\dproj$; dashed
      polylines), $T_X$ (dotted), or both (solid). The likelihood is
      evaluated at available snapshots and interpolated linearly
      between them. Early times are excluded by the relic, and late
      times by the XSB morphology. In the intermediate region, the
      interplay of $T_X$ and kinematics leaves a dominant peak around
      400 Myr for either impact parameter. A single plot works for
      both $b=0.5$ and 1 Mpc because the rectangular exclusion
      zones are quite similar for both cases. In contrast, $b=0$ is not
      shown because it is excluded everywhere by its XSB morphology
      and/or relic position.}}
\label{fig-timeline}
\end{figure}

\rev{Even so, we marginalize the kinematic constraint over LOS to
  present a summary of TSP constraints in
  Figure~\ref{fig-timeline}. Although no single line of evidence is
  highly constraining, combining the various lines of evidence leaves
  a single dominant likelihood peak at TSP${\approx}400$ Myr.}

\rev{\textit{Caveats.}} We caution that these numbers were obtained via
comparison with a limited suite of simulations \rev{in which initial
  velocity and initial ICM temperature were not varied.}  Varying the
initial velocity would vary the relationships between subcluster
separation, velocity, and TSP.  \rev{Varying the initial ICM
  temperature would affect $T_X$ even after pericenter.  Therefore,
  the data could be a good fit to additional models not presented
  here, perhaps at different TSP and viewed along different LOS. In
  other words, we have established only that the data are consistent
  with a specific subset of readily available public models.  We
  present this as a starting point for additional modeling rather than
  as a set of conclusive constraints.}

\textit{Implications for optical selection of merging clusters.}  The
redMaPPer BCG candidates in Figure~\ref{fig-relicsACS} illustrate
some pitfalls regarding the selection of bimodal cluster candidates
using redMaPPer BCGs alone.  The BCG candidate found most likely by
redMaPPer (labeled 0 in Figure~\ref{fig-relicsACS}) is on the southern
periphery of the southern cluster and does not seem to be germane to
the merger.  Furthermore, BCG candidates may be a function of angular
resolution. BCG 0 is brightest in SDSS photometry in large part
because it is a blend of three galaxies, as revealed by the ACS
imaging. It is possible that higher angular resolution imaging would
have prevented this peripheral galaxy from becoming a top BCG
candidate. However, BCG 1, at the southern X-ray peak, is also blended
in SDSS photometry, hence it is not clear that this would become the
top BCG candidate at higher angular resolution. Algorithms using all
potential photometric members \citep[e.g.,][]{WHY2024} may prove to be
more robust.

\textit{Similarity to a well-studied merger.}  Abell 115 is another
binary merger in which two distinct XSB peaks yield a separation
vector that does not point to the relic
\citep{Botteon2016A115,Kim2019A115,Wonki2020A115}.  In that system,
much deeper X-ray data (361 ks with \textit{Chandra}) revealed that
each subcluster has ICM tails pointing nearly perpendicularly to the
current separation vector.  In a detailed comparison with hydrodynamic
simulations, \citet{Wonki2020A115} found a substantial impact
parameter and an observation time ``${\sim}$0.3 Gyr after the impact,
before the two subclusters reach their apocenters,'' which is similar
to our findings for RXC J0032.1+1808.  Some differences between the
systems should also be noted. Abell 115 has a less equal mass ratio
\citep[2:1;][]{Kim2019A115} and is much less optically rich \citep[113
vs 248;][]{Rykoff2016}. Abell 115 has more striking radio and X-ray
morphology with a larger relic and clear ICM tails, but the salience
of these details may stem partly from the higher-quality data and
lower redshift (at $z=0.197$, less than half the luminosity distance
of RXC J0032.1+1808). The mass comparison is uncertain: the optical
richness relation predicts that RXC J0032.1+1808 has nearly triple the
mass of Abell 115, but the SZE relation predicts that they are equal
\citep{PSZ2}. A wide-field lensing study that captures the total mass
of RXC J0032.1+1808, as \citet{Kim2019A115} did for Abell 115, would
clarify the comparison.

Abell 115 may have a larger impact parameter as
follows. \citet{Wonki2020A115} found a most likely pericenter distance
of 500 kpc for Abell 115.  We find that the $b{=}0.5$ (1.0) Mpc run
from \citet{ZuHone11} produces a 109 (237) kpc pericenter distance,
hence we surmise that Abell 115 would be incompatible with these
models while RXC J0032.1+1808 is compatible.  

\textit{Future directions.} To make further progress on RXC
J0032.1+1808, deeper X-ray data enabling temperature maps and shock
detection would be useful.  A wider-field lensing study that captures
the total mass would fix the overall mass calibration, which we are
currently estimating from scaling relations.  In the radio, a spectral
index map and/or polarization measurements of the relic would further
increase confidence that its emission traces a merger shock; its
classification in \citet{Botteon2022} rests on morphology and lack of
association with an AGN. Polarization measurements could also help
constrain the viewing angle.  At visible wavelengths, measuring the
subcluster relative velocity via member galaxies more precisely would
tighten the constraints on TSP and viewing angle\rev{---but would
  require a much more extensive redshift survey, particularly to
  understand the relationship between the BCG velocities and those of
  the other members}.  Making optimal use of such rich
  datasets would require more detailed exploration of simulations,
  with more varied initial conditions (or an ensemble of analogs from
  cosmological simulations) and projections along arbitrary LOS.

An additional observable not considered in the work is the offset
between XSB and mass/galaxy peaks visible in
Figure~\ref{fig-relicsACS}.  In the `ram pressure slingshot'
\citep{Hallman04}, ram pressure sets the gas back from the mass peak
initially; the gas then falls back to, and \textit{past}, the center
of the potential.  The presence and sign of the offset is therefore a
potentially useful diagnostic of TSP. However, the apparent offset
would vary with LOS such that  joint modeling of all observables along
arbitrary LOS would be required.

\acknowledgments This work was supported by NSF grant number 2308383.
Some of the data presented herein were obtained at Keck Observatory,
which is a private 501(c)3 non-profit organization operated as a
scientific partnership among the California Institute of Technology,
the University of California, and the National Aeronautics and Space
Administration. The Observatory was made possible by the generous
financial support of the W. M. Keck Foundation.  The authors wish to
recognize and acknowledge the very significant cultural role and
reverence that the summit of Maunakea has always had within the Native
Hawaiian community. We are most fortunate to have the opportunity to
conduct observations from this mountain.  

This work used data from the Legacy Surveys. The Legacy Surveys
consist of three individual and complementary projects: the Dark
Energy Camera Legacy Survey (DECaLS; Proposal ID \#2014B-0404; PIs:
David Schlegel and Arjun Dey), the Beijing-Arizona Sky Survey (BASS;
NOAO Prop. ID \#2015A-0801; PIs: Zhou Xu and Xiaohui Fan), and the
Mayall z-band Legacy Survey (MzLS; Prop. ID \#2016A-0453; PI: Arjun
Dey). DECaLS, BASS and MzLS together include data obtained,
respectively, at the Blanco telescope, Cerro Tololo Inter-American
Observatory, NSF’s NOIRLab; the Bok telescope, Steward Observatory,
University of Arizona; and the Mayall telescope, Kitt Peak National
Observatory, NOIRLab. Pipeline processing and analyses of the data
were supported by NOIRLab and the Lawrence Berkeley National
Laboratory (LBNL). The Legacy Surveys project is honored to be
permitted to conduct astronomical research on Iolkam Du’ag (Kitt
Peak), a mountain with particular significance to the Tohono O’odham
Nation.

NOIRLab is operated by the Association of Universities for Research in
Astronomy (AURA) under a cooperative agreement with the National
Science Foundation. LBNL is managed by the Regents of the University
of California under contract to the U.S. Department of Energy.

This project used data obtained with the Dark Energy Camera (DECam),
which was constructed by the Dark Energy Survey (DES)
collaboration. Funding for the DES Projects has been provided by the
U.S. Department of Energy, the U.S. National Science Foundation, the
Ministry of Science and Education of Spain, the Science and Technology
Facilities Council of the United Kingdom, the Higher Education Funding
Council for England, the National Center for Supercomputing
Applications at the University of Illinois at Urbana-Champaign, the
Kavli Institute of Cosmological Physics at the University of Chicago,
Center for Cosmology and Astro-Particle Physics at the Ohio State
University, the Mitchell Institute for Fundamental Physics and
Astronomy at Texas A\&M University, Financiadora de Estudos e
Projetos, Fundacao Carlos Chagas Filho de Amparo, Financiadora de
Estudos e Projetos, Fundacao Carlos Chagas Filho de Amparo a Pesquisa
do Estado do Rio de Janeiro, Conselho Nacional de Desenvolvimento
Cientifico e Tecnologico and the Ministerio da Ciencia, Tecnologia e
Inovacao, the Deutsche Forschungsgemeinschaft and the Collaborating
Institutions in the Dark Energy Survey. The Collaborating Institutions
are Argonne National Laboratory, the University of California at Santa
Cruz, the University of Cambridge, Centro de Investigaciones
Energeticas, Medioambientales y Tecnologicas-Madrid, the University of
Chicago, University College London, the DES-Brazil Consortium, the
University of Edinburgh, the Eidgenossische Technische Hochschule
(ETH) Zurich, Fermi National Accelerator Laboratory, the University of
Illinois at Urbana-Champaign, the Institut de Ciencies de l’Espai
(IEEC/CSIC), the Institut de Fisica d’Altes Energies, Lawrence
Berkeley National Laboratory, the Ludwig Maximilians Universitat
Munchen and the associated Excellence Cluster Universe, the University
of Michigan, NSF’s NOIRLab, the University of Nottingham, the Ohio
State University, the University of Pennsylvania, the University of
Portsmouth, SLAC National Accelerator Laboratory, Stanford University,
the University of Sussex, and Texas A\&M University.

BASS is a key project of the Telescope Access Program (TAP), which has
been funded by the National Astronomical Observatories of China, the
Chinese Academy of Sciences (the Strategic Priority Research Program
“The Emergence of Cosmological Structures” Grant \# XDB09000000), and
the Special Fund for Astronomy from the Ministry of Finance. The BASS
is also supported by the External Cooperation Program of Chinese
Academy of Sciences (Grant \# 114A11KYSB20160057), and Chinese National
Natural Science Foundation (Grant \# 12120101003, \# 11433005).

The Legacy Survey team makes use of data products from the Near-Earth
Object Wide-field Infrared Survey Explorer (NEOWISE), which is a
project of the Jet Propulsion Laboratory/California Institute of
Technology. NEOWISE is funded by the National Aeronautics and Space
Administration.

The Legacy Surveys imaging of the DESI footprint is supported by the
Director, Office of Science, Office of High Energy Physics of the
U.S. Department of Energy under Contract No. DE-AC02-05CH1123, by the
National Energy Research Scientific Computing Center, a DOE Office of
Science User Facility under the same contract; and by the
U.S. National Science Foundation, Division of Astronomical Sciences
under Contract No. AST-0950945 to NOAO.

\facilities{Keck:II (DEIMOS), XMM} 

\software{SAS (v19.0.0; Gabriel et al. 2004), PypeIt (Prochaska et
  al. 2020)}

\vskip5cm 
\bibliography{ms}
\end{document}

%% file: ztable.tex
7.965975 & 17.995422 & 0.139151 & 0.000038\\
7.979279 & 17.991050 & 0.375415 & 0.000028\\
7.994463 & 18.040633 & 0.611399 & 0.000114\\
7.998640 & 17.994210 & 0.370238 & 0.000047\\
8.002850 & 18.066533 & 0.371189 & 0.000038\\
8.003279 & 18.187167 & 0.364485 & 0.000176\\
8.005467 & 18.122794 & 0.375075 & 0.000082\\
8.009083 & 18.025069 & 0.250609 & 0.000085\\
8.009329 & 18.048553 & 0.373407 & 0.000091\\
8.011833 & 17.956300 & 0.127827 & 0.000034\\
8.013879 & 18.055311 & 0.373891 & 0.000050\\
8.017896 & 18.169400 & 0.364952 & 0.000018\\
8.020617 & 18.179881 & 0.479828 & 0.000049\\
8.021817 & 18.090786 & 0.374958 & 0.000062\\
8.023483 & 18.039017 & 0.366269 & 0.000051\\
8.027679 & 18.109256 & 0.374224 & 0.000098\\
8.027738 & 17.970894 & 0.251393 & 0.000010\\
8.028975 & 18.100478 & 0.375559 & 0.000103\\
8.030029 & 17.949767 & 0.253010 & 0.000070\\
8.031367 & 18.110900 & 0.376943 & 0.000087\\
8.033971 & 18.048797 & 0.372023 & 0.000101\\
8.034562 & 18.124622 & 0.368220 & 0.000216\\
8.035442 & 18.024236 & 0.370539 & 0.000027\\
8.035896 & 18.026458 & 0.142854 & 0.000009\\
8.037433 & 18.067414 & 0.377944 & 0.000101\\
8.039121 & 18.151392 & 0.366469 & 0.000067\\
8.039687 & 18.115253 & 0.380412 & 0.000132\\
8.041554 & 18.045786 & 0.367136 & 0.000103\\
8.043204 & 18.130033 & 0.375208 & 0.000078\\
8.043271 & 18.005903 & 0.367537 & 0.000111\\
8.047025 & 18.118947 & 0.377160 & 0.000060\\
8.047521 & 18.054528 & 0.375158 & 0.000153\\
8.047688 & 18.140936 & 0.367020 & 0.000057\\
8.049150 & 18.132675 & 0.378094 & 0.000351\\
8.049858 & 18.138519 & 0.379428 & 0.000267\\
8.051913 & 18.144917 & 0.370472 & 0.000102\\
8.055212 & 18.098456 & 0.377093 & 0.000135\\
8.057200 & 18.071239 & 0.364351 & 0.000245\\
8.061933 & 18.168200 & 0.374408 & 0.000134\\
8.062429 & 18.109708 & 0.380762 & 0.000096\\
8.067021 & 18.159397 & 0.372340 & 0.000067\\
8.069000 & 18.201060 & 0.143387 & 0.000042\\
8.080321 & 18.177294 & 0.138667 & 0.000021\\